\documentclass[11pt]{article}

\usepackage[preprint]{acl}

\usepackage{times}
\usepackage{latexsym}

\usepackage[T1]{fontenc}

\usepackage[utf8]{inputenc}

\usepackage{microtype}

\usepackage{inconsolata}

\usepackage{graphicx}

\usepackage{booktabs}
\usepackage{multirow}
\usepackage{amsmath}
\usepackage{amssymb}
\usepackage{xcolor}    
\usepackage{colortbl}
\definecolor{goodgreen}{RGB}{0, 150, 0}
\newcommand{\inc}[1]{\textcolor{goodgreen}{\scriptsize{(+#1)}}}
\usepackage{enumitem}

\title{FailureMem: A Failure-Aware Multimodal Framework for Autonomous Software Repair}

\author{
  \textbf{Ruize Ma}\textsuperscript{1,2,}\thanks{Equal contribution. $\dagger$ Corresponding author. Email: \texttt{fy@nju.edu.cn}},
  \textbf{Yilei Jiang}\textsuperscript{3,$*$},
  \textbf{Shilin Zhang}\textsuperscript{1,2,$*$},
  \textbf{Zheng Ma}\textsuperscript{2},
  \textbf{Yi Feng}\textsuperscript{1,$\dagger$}, \\
  \textbf{Vincent Ng}\textsuperscript{4},
  \textbf{Zhi Wang}\textsuperscript{1},
  \textbf{Xiangyu Yue}\textsuperscript{3},
  \textbf{Chuanyi Li}\textsuperscript{1},
  \textbf{Lewei Lu}\textsuperscript{2} \\
  \\
  \textsuperscript{1}Nanjing University \quad
  \textsuperscript{2}SenseTime \quad
  \textsuperscript{3}The Chinese University of Hong Kong \\
  \textsuperscript{4}University of Texas at Dallas
}

\usepackage{caption}

\begin{document}
\maketitle

\begin{abstract}

Multimodal Automated Program Repair (MAPR) extends traditional program repair by requiring models to jointly reason over source code, textual issue descriptions, and visual artifacts such as GUI screenshots. While recent LLM-based repair systems have shown promising results, existing approaches face several limitations: rigid workflow pipelines restrict exploration during debugging, visual reasoning is often performed over full-page screenshots without localized grounding, and failed repair attempts are rarely transformed into reusable knowledge.
To address these challenges, we propose FailureMem, a multimodal repair framework that integrates three key mechanisms: a hybrid workflow–agent architecture that balances structured localization with flexible reasoning, active perception tools that enable region-level visual grounding, and a Failure Memory Bank that converts past repair attempts into reusable guidance. 
Experiments on SWE-bench Multimodal demonstrate FailureMem improves the resolved rate over GUIRepair by 3.7\%.Our code is available at \url{https://github.com/Ruize-Ma/FailureMem}.

\end{abstract}

\section{Introduction}
\label{sec:intro}
\begin{figure*}[t]
\centering
\includegraphics[width=\textwidth]{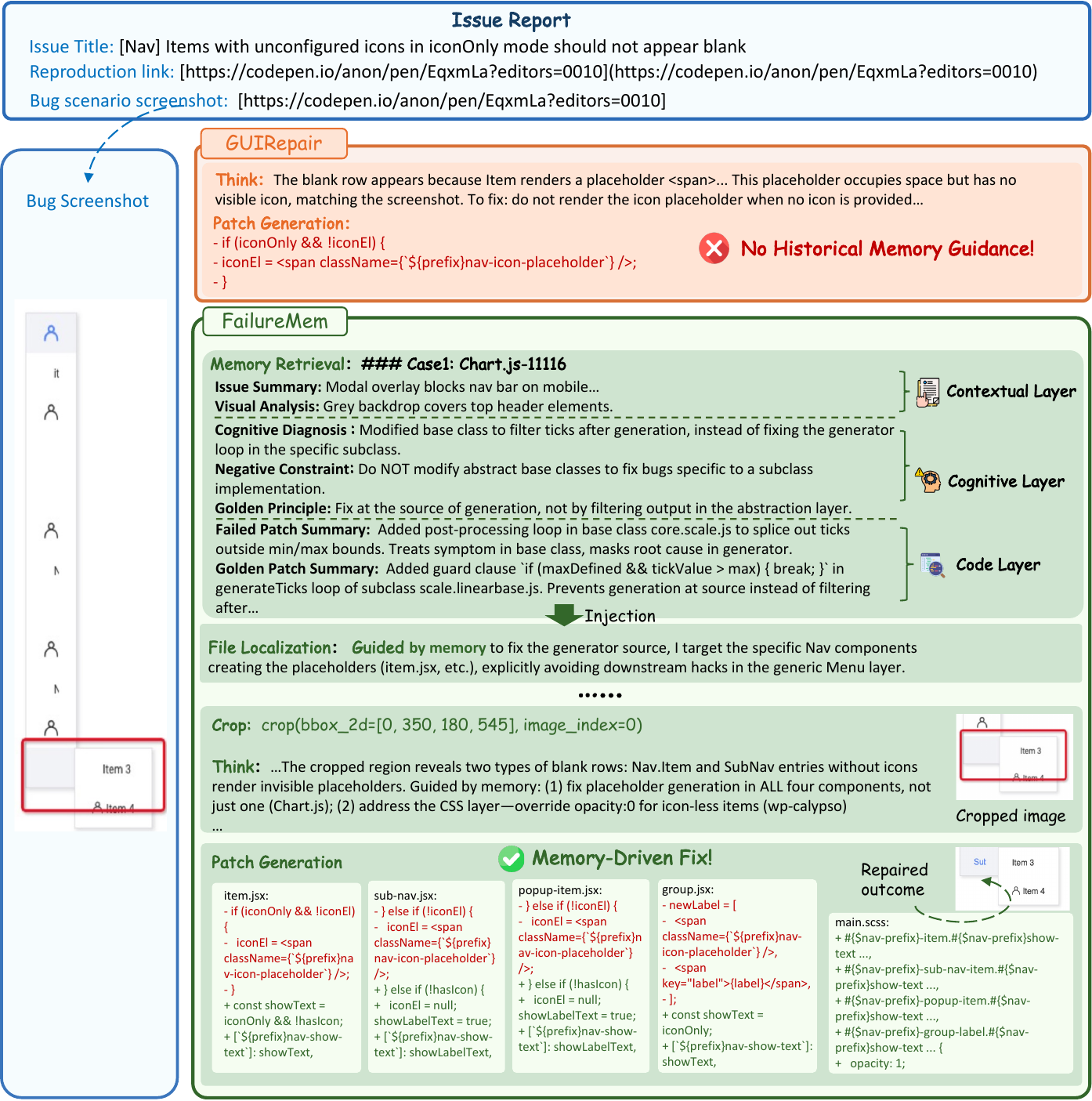}
\caption{GUIRepair vs. FailureMem on \texttt{alibaba-fusion/next-895}. FailureMem retrieves historical repair guidance, crops the screenshot to isolate the bug region, and produces a comprehensive five-file patch.}
\label{fig:teaser}
 \vspace{-2mm}
\end{figure*}

Software defects are an inevitable part of software development. Automated Program Repair (APR) aims to automatically generate patches that fix buggy programs while preserving intended behavior, typically using signals such as source code and test cases to synthesize code modifications~\cite{LLM4APR_Huang_ASE,LLM4APR_Huang_TSE,LLM4APR_Jiang,LLM4APR_Wu,NTR}.

Recent advances in large language models (LLMs) have significantly expanded the capability of APR systems. Leveraging their strong code understanding and generation abilities, LLM-based approaches have progressed from function-level repair via fine-tuning and prompting~\cite{LLM4APR_Huang_ASE,LLM4APR_Jiang,AlphaRepair,ChatRepair} to repository-scale issue resolution, where autonomous agents analyze entire codebases and generate patches for real-world software issues~\cite{SWE-agent,Agentless,AutoCodeRover}. Benchmarks such as SWE-bench have been proposed to evaluate this capability by requiring models to resolve GitHub issues through patch generation that passes repository test suites.

However, real-world debugging often relies on information beyond code and text. Many issue reports include visual artifacts, such as screenshots of graphical user interfaces, diagrams, or UI mockups, which convey important clues about the intended system behavior. To capture this richer context, recent work introduced SWE-bench Multimodal, which augments issue instances with visual information, giving rise to the task of Multimodal Automated Program Repair (MAPR). Compared with text-only settings, solving such tasks requires models to jointly reason over code, natural language descriptions, and visual observations, making the repair problem substantially more challenging.

Formally, MAPR can be defined as follows: given a repository, a problem description (e.g., a GitHub issue), and associated visual artifacts (e.g., screenshots), the goal is to generate a patch that modifies the repository to resolve the reported issue. Figure~\ref{fig:teaser} illustrates a representative instance. The issue report describes a rendering bug in a Nav component and includes a screenshot showing the faulty sidebar; the task is to generate a patch that fixes the visual defect reflected in the interface. Existing methods, such as GUIRepair~\cite{guirepair}, have taken an important step toward addressing this problem by incorporating visual inputs into the repair process. However, current approaches still face several practical challenges.

\vspace{-2mm}
\paragraph{Challenge 1.} Existing methods struggle to balance structured reasoning and exploratory search. Workflow-based approaches follow predefined pipelines (e.g., issue understanding, file localization, patch generation), which efficiently constrain the search space but may overlook alternative hypotheses. In contrast, agentic systems enable flexible exploration via autonomous reasoning, yet they may lose direction during multi-step exploration and repeatedly inspect irrelevant components.


\vspace{-2mm}
\paragraph{Challenge 2.} Visual reasoning is often performed over full-page screenshots, which distribute attention across many irrelevant interface regions. In Figure~\ref{fig:teaser}, the screenshot contains multiple UI elements, including the main content area, navigation links, and surrounding layout components, while the actual defect is localized to the sidebar rendering of the Nav component. Processing the entire screenshot equally can dilute the model’s attention and make it harder to isolate the precise visual anomaly that should guide the repair process.
\vspace{-2mm}
\paragraph{Challenge 3.} Current repair agents rarely transform previous repair attempts into reusable knowledge. For instance, when attempting to fix the sidebar rendering issue, a repair agent may generate several candidate patches that modify layout properties or component structure. However, these attempts are typically treated as isolated trials rather than structured experiences. As a result, the agent cannot effectively retrieve analogous repair cases, analyze why earlier fixes failed (e.g., modifying the wrong layout container), or apply concrete repair pattern, such as adjusting component-level layout constraints, to guide subsequent repair attempts.

To address these challenges, we propose \textbf{FailureMem}, a multimodal repair framework for GUI issue resolution. FailureMem combines three key designs. First, it introduces \textbf{a hybrid workflow–agent architecture} that keeps file and element localization within a structured workflow while reserving an agentic loop for patch generation, enabling controlled search with flexible exploration. Second, it improves visual grounding through \textbf{active perception tools}, including \textit{Crop} and \textit{Grounding}, together with an interactive \textit{Bash} environment that allows the agent to inspect repositories and verify assumptions before editing code. Third, it incorporates a \textbf{Failure Memory Bank} that transforms historical repair trajectories into reusable guidance via contextual, cognitive, and code-level representations. We evaluate FailureMem on SWE-bench Multimodal. With GPT-5.1, it improves the resolved rate over GUIRepair (a SOTA MAPR model) by 3.7\%.


Our contributions are four-fold: (1) We propose a hybrid workflow–agent architecture for multimodal program repair that balances structured localization with flexible patch exploration; (2) We introduce active perception tools for localized visual grounding together with an interactive repository exploration environment; (3) We design a Failure Memory Bank that converts historical repair trajectories into reusable guidance for both localization and patch generation; (4) Experiments on SWE-bench Multimodal demonstrate consistent improvements over strong baselines.

\begin{figure*}[t]
    \centering
    \includegraphics[width=\textwidth]{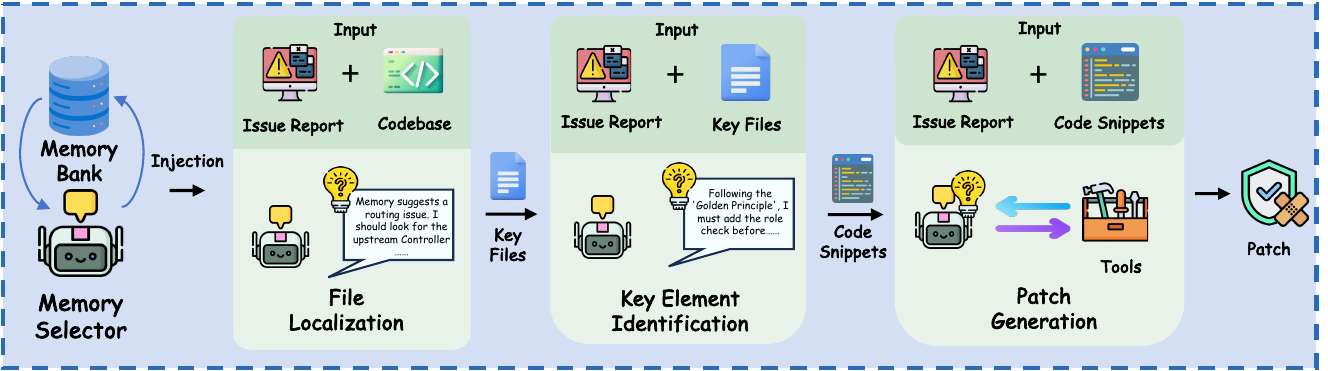}
    \caption{Overview of FailureMem. 
    }
    \label{fig:main_framework}
    \vspace{-2mm}
\end{figure*}

\section{A Motivating Example}
\label{sec:motivation}

We use an example from SWE-bench Multimodal (Figure~\ref{fig:teaser}) to illustrate the challenges of MAPR and preview how FailureMem addresses them.

Issue \texttt{alibaba-fusion/next-895} reports that in the Nav component's \texttt{iconOnly} mode, items without an \texttt{icon} prop render as blank rows in the vertical sidebar. The screenshot shows icon entries interspersed with empty horizontal bars.
\vspace{-2mm}
\paragraph{How GUIRepair Fails.}
GUIRepair localizes two files (\texttt{item.jsx} and \texttt{nav.jsx}) but misses other affected components and the Nav stylesheet. Its patch modifies only \texttt{item.jsx}, removing the placeholder logic for \texttt{Nav.Item}. However, the bug spans two layers: four components generate invisible placeholders (JS layer), and a global CSS rule hides labels under \texttt{iconOnly} mode (CSS layer). The patch therefore fixes one component in one layer and fails regression tests.

This failure reflects the three challenges from Section~\ref{sec:intro}.
First, GUIRepair follows a fixed localization workflow that restricts exploration, causing it to miss the broader set of affected components and style rules.
Second, visual reasoning operates on the full-page screenshot without region-level focus, making it difficult to associate the visual symptom (empty rows between icons) with placeholder elements generated by multiple Nav components.
Third, repair attempts are treated as isolated trials, preventing the system from leveraging historical repair experience or reusable repair patterns.
\vspace{-2mm}
\paragraph{How FailureMem Works.}
FailureMem first retrieves historical cases from a Failure Memory Bank (Challenge 3). Two cases provide useful guidance.
A Chart.js axis-tick bug emphasizes fixing errors at the source of generation rather than filtering outputs in abstraction layers.
A wp-calypso layout bug advises targeted CSS adjustments instead of modifying reusable component logic.

Guided by these principles, the model localizes all four affected components and the Nav stylesheet, going beyond the two files identified by GUIRepair. Unlike GUIRepair’s rigid workflow—where early localization errors propagate through the pipeline, FailureMem combines structured workflow with agentic reasoning, allowing the agent to revisit earlier steps and expand the search when inconsistencies arise (Challenge 1).

During patch generation, the agent uses visual cropping to isolate the sidebar region, confirming that blank rows correspond to icon-less entries rendering invisible placeholders (Challenge 2). The resulting patch modifies five files: it removes placeholder generation from four components (\texttt{Item}, \texttt{SubNav}, \texttt{PopupItem}, \texttt{Group}), introduces a \texttt{nav-show-text} CSS class for icon-less items, and adds an \texttt{opacity:\,1} override in \texttt{main.scss}. The previously blank rows now display text labels, and the patch passes all tests.

\section{Method}


We propose FailureMem, a MAPR framework combining a hybrid workflow–agent architecture, active perception tools, and a failure memory bank. Figure~\ref{fig:main_framework} shows the overall architecture.

\subsection{Problem Formulation}
Given a software repository with codebase $\mathcal{C}$ and a multimodal issue report comprising a textual description $\mathcal{D}$ and visual screenshots $\mathcal{I}$, MAPR aims at generating a code patch $\Delta$ such that the patched codebase $\mathcal{C}' = \mathcal{C} \oplus \Delta$ passes a held-out test suite:
\begin{equation}
\mathcal{F}(\mathcal{C}') \models T_{spec}
\end{equation}
where $\oplus$ is the patch application operator, 
$\mathcal{F}(\cdot)$ denotes program execution, and $T_{spec}$ includes both fail-to-pass tests (verifying the issue is fixed) and pass-to-pass tests (verifying no regression). 

\subsection{Failure-aware Retrieval Module}
\label{sec:failuremem}


LLM-based repair agents often produce plausible but incorrect patches and may repeat errors from prior attempts. To address this limitation, we introduce a failure memory bank, which stores historical repair trajectories. We further design a failure-aware retrieval module that contrasts failed patches with ground-truth fixes to extract reusable repair patterns. By retrieving and reusing prior experiences, the model reduces recurring mistakes and improves transfer to new issues.
\vspace{-2mm}
\paragraph{Hierarchical Memory Structure Design.}
The failure memory bank is defined as a collection of $N$ entries: $\mathcal{B} = \{\mathcal{M}_i\}_{i=1}^{N}$. 
As illustrated in Figure~\ref{fig:teaser}, we design three complementary layers to capture different levels of repair knowledge in each entry.
$$\mathcal{M}i = \langle \mathcal{L}{ctx}^{(i)},\ \mathcal{L}{cog}^{(i)},\ \mathcal{L}{code}^{(i)} \rangle$$

Contextual Layer ($\mathcal{L}_{ctx}$) 
provides retrieval keys for matching similar cases. It contains two text-based fields: the \textit{Issue Summary}, which abstracts the bug scenario, and the \textit{Visual Analysis}, which describes observable visual symptoms (e.g., ``The modal overlay obscures the navigation bar''). Instead of storing raw screenshots, we encode visual information as text. This decision reduces token cost during retrieval and avoids background noise in images, ensuring that the selector focuses on bug-relevant signals.




Cognitive Layer ($\mathcal{L}_{cog}$) provides high-level reasoning guidance. It includes a \textit{Cognitive Diagnosis} that explains the causal mechanism of failure, a \textit{Negative Constraint} that prohibits incorrect repair strategies (e.g., ``Do not modify downstream rendering logic for upstream data errors''), and a \textit{Golden Principle} that captures transferable design patterns (e.g., fixing errors at the source of generation rather than applying downstream filtering). This layer focuses on abstract repair knowledge to improve generalization across issues.

In contrast, Code Layer ($\mathcal{L}_{code}$) supplies implementation-level evidence. It stores \textit{Failed} and \textit{Golden Patch Summaries}, explicitly highlighting structural divergences between incorrect and correct solutions. By grounding abstract principles in concrete code changes, this layer supports actionable and executable repairs.
Together, the Cognitive Layer shapes repair strategy at the reasoning level, while the Code Layer ensures alignment with precise implementation patterns.




\vspace{-2mm}
\paragraph{Failure Memory Bank Construction.}
To construct the memory bank, we process historical failure cases offline (see Appendix~\ref{sec:app_data_construction} for detailed data statistics and the distillation pipeline). 
For each case, we collect the agent’s failed patch $P_{fail}$, the developer’s ground-truth patch $P_{gold}$, and the original issue report, including the textual description $\mathcal{D}$ and screenshots $\mathcal{V}$. Using Gemini 3 Pro as the backbone model, we analyze these multimodal inputs and distill them into hierarchical memory entries.
Formally, for each historical failure case $i$, the construction process is:
$$\mathcal{M}i = f{\text{distill}}(P_{fail}^{(i)},\ P_{gold}^{(i)},\ \mathcal{D}^{(i)},\ \mathcal{V}^{(i)})$$
where $f_{\text{distill}}$ denotes the distillation model that takes the failed patch, ground-truth patch, issue description, and visual screenshots as inputs, and outputs the structured three-layer memory entry.

The model is prompted to generate structured content for all three memory layers. For the Contextual Layer, it abstracts the issue report into a concise \textit{Issue Summary} and converts screenshots $\mathcal{V}$ into a textual \textit{Visual Analysis}, preserving visual symptoms for efficient retrieval. For the Cognitive Layer, it contrasts $P_{fail}$ and $P_{gold}$ to identify the root cause, extracting the \textit{Cognitive Diagnosis}, \textit{Negative Constraints}, and \textit{Golden Principles}. For the Code Layer, it summarizes the implementation differences into \textit{Failed} and \textit{Golden Patch Summaries}, capturing key divergences while filtering out project-specific noise.

\vspace{-2mm}
\paragraph{Retrieving Memory Entries.}
Before file localization, we retrieve relevant historical memory entries. We employ a Selector Agent to identify the top-$k$ most relevant cases.
Given a new issue with description $\mathcal{D}_q$ and screenshots $\mathcal{V}_q$, the Selector Agent receives all candidate Contextual Layers and directly selects the top-$k$ most relevant entries in a single pass:
$$\mathcal{R} = \mathrm{Selector}\!\left(\mathcal{D}_q,\ \mathcal{V}_q,\ \{\mathcal{L}_{ctx}^{(i)}\}_{i=1}^{N},\ k\right)$$
where $\mathcal{R} \subset \mathcal{B}$ and $|\mathcal{R}| = k$.

The retrieval process is modeled as a semantic selection task. The Selector Agent is presented with the current issue report, which contains both the textual description $\mathcal{D}$ and the associated visual screenshots $\mathcal{V}$. It compares these inputs against the \textit{Contextual Layer} ($\mathcal{L}_{ctx}$) of the candidate memories, specifically aligning the current problem with the historical \textit{Issue Summary} and \textit{Visual Analysis}. This approach allows the agent to leverage its semantic understanding to find relevant precedents based on symptom similarity, rather than relying on superficial keyword overlap.

Once $\mathcal{R}$ is obtained, the guidance context is constructed by extracting the Cognitive and Code layers from each retrieved entry:
$$\mathcal{G} = \{(\mathcal{L}{cog}^{(i)},\ \mathcal{L}{code}^{(i)}) \mid \mathcal{M}i \in \mathcal{R}\}$$
This guidance $\mathcal{G}$ is then injected into the model's context at each subsequent phase.

\subsection{Repair with a Hybrid Agent Framework}
\label{sec:hybrid_arch}
After retrieving the top-$k$ memory entries, we adopt a hybrid agent–workflow architecture that balances structured fault localization with flexible patch generation. To reduce computational overhead in repository-scale contexts, we confine the agentic loop to the final patch generation stage, while using a deterministic workflow for earlier localization steps. This design narrows the search space and improves token efficiency. At each phase, the injected Cognitive and Code layers further align the model’s reasoning with historical constraints, guiding both localization and repair decisions.
\vspace{-2mm}
\paragraph{Phase 1: File Localization.}
The process begins by scanning the repository's directory tree to identify a candidate set of files relevant to the issue. The model receives the issue description, the file tree structure, and the retrieved memory entries. At this stage, the injected \textit{Cognitive Layer} helps the model filter out irrelevant domains (e.g., distinguishing between test files and source code) based on historical failure patterns. This phase operates in a single-pass inference mode to efficiently narrow the search space without invoking external tools.
\vspace{-2mm}
\paragraph{Phase 2: Key Element Identification.}
Once the candidate files are determined, the system identifies the specific classes or functions requiring modification. To handle the token limit, we utilize a \textit{Skeleton Compression} strategy (detailed in Appendix~\ref{app:skeleton}). We parse the candidate files to retain only class signatures, function headers, and docstrings, while abstracting implementation bodies. The model analyzes these skeletons alongside the injected memory to pinpoint the key elements. The memory injection here is critical for preventing architectural misalignments, such as guiding the model to focus on the controller logic rather than view definitions when the memory suggests a backend root cause. Like the previous phase, this step does not utilize an agentic loop.
\vspace{-2mm}
\paragraph{Phase 3: Agentic Patch Generation.}
This final stage is the only phase where the model operates as an autonomous agent with a multi-turn execution loop. The input comprises the \textit{full raw code} of the identified key elements, the issue description, and the retrieved memory. We equip the agent with a specific suite of tools to verify logic and resolve multimodal ambiguities:

(1) \textit{Active Visual Perception.} To address the resolution limitations of standard multimodal models, the agent utilizes a \textit{Crop} tool to zoom into specific regions for detailed inspection. Additionally, a \textit{Grounding} tool enables the agent to draw bounding boxes around bug-related UI elements. This action explicitly focuses the model's attention on relevant visual contexts, mitigating the noise from unrelated interface components. Implementation details are provided in Appendix~\ref{app:perception}.

(2) \textit{Interactive Environment.} A \textit{Bash} tool (operating within a strictly sandboxed execution environment as described in Appendix~\ref{app:bash}) allows the agent to execute commands. This enables the agent to actively explore the repository structure and verify code logic dynamically, such as checking dependency versions or running reproduction scripts, before committing to specific modifications.

During this loop, the model iteratively refines its plan using these tools, while continuously receiving guidance from the injected \textit{Cognitive} and \textit{Code} layers. These memory layers help the agent apply proven repair principles, reference successful implementation patterns, and avoid failure modes identified in past repair attempts. This dual-sided memory guidance significantly improves the agent's ability to make reliable repair decisions.

\section{Experiment}

\begin{table*}[t]
    \centering

    \resizebox{\textwidth}{!}{
        \setlength{\tabcolsep}{3.5pt} 
        \begin{tabular}{ll c | cccccccccccc}
            \toprule
            
            \multirow{2}{*}{\textbf{Method}} & \multirow{2}{*}{\textbf{LLM}} &
            \multirow{2}{*}{\textbf{(Resolved Rate \%)}} &
            \multicolumn{12}{c}{\textbf{Repository Breakdown (Resolved Rate \%)}} \\
            \cmidrule(l){4-15}

            & & &
            \rotatebox{90}{next} & \rotatebox{90}{bpmn-js} & \rotatebox{90}{carbon} & \rotatebox{90}{eslint} &
            \rotatebox{90}{lighthouse} & \rotatebox{90}{grommet} & \rotatebox{90}{highlight.js} & \rotatebox{90}{openlayers} &
            \rotatebox{90}{prettier} & \rotatebox{90}{prism} & \rotatebox{90}{quarto-cli} & \rotatebox{90}{scratch-gui} \\

            \midrule
            
            \multicolumn{15}{l}{\textbf{Other Baselines}} \\
            SWE-agent Multimodal & GPT-4o & 12.4 & 0.0 & 27.8 & 1.5 & 0.0 & 5.6 & 0.0 & 2.6 & 51.9 & 7.7 & 0.0 & 0.0 & 0.0 \\
            Computer-Use Agents & GPT-4o & 20.1 & - & - & - & - & - & - & - & - & - & - & - & - \\
            Agentless Lite & GPT-5.1 & 28.4 & 15.4 & 53.7 & 14.9 & 9.1 & 9.3 & 4.8 & 10.3 & 94.9 & 7.7 & 2.6 & 0.0 & 0.0 \\
            Zencoder & NA & 27.5 & 15.4 & 51.9 & 11.2 & 18.2 & 13.0 & 19.0 & 10.3 & 81.0 & 7.7 & 18.4 & 8.3 & 0.0 \\

            \midrule
            \multicolumn{15}{l}{\textbf{GUIRepair vs. Ours (GPT-4.1)}} \\
            GUIRepair & GPT-4.1 & 28.8 & 17.9 & 66.7 & 12.7 & 36.4 & 7.4 & 4.8 & 2.6 & 96.2 & 7.7 & 2.6 & 4.2 & 0.0 \\
            \rowcolor{gray!10}
            \textbf{FailureMem (Ours)} & GPT-4.1 & \textbf{31.1} \inc{2.3} & 23.1 & 68.5 & 12.7 & 18.2 & 13.0 & 9.5 & 5.1 & 97.5 & 15.4 & 13.2 & 4.2 & 0.0 \\
            \midrule

            \multicolumn{15}{l}{\textbf{GUIRepair vs. Ours (GPT-5.1)}} \\
            GUIRepair & GPT-5.1 & 29.4 & 23.1 & 57.4 & 11.9 & 36.4 & 13.0 & 4.8 & 5.1 & 97.5 & 15.4 & 5.3 & 4.2 & 0.0 \\

            \rowcolor{gray!10}
            \textbf{FailureMem (Ours)} & GPT-5.1 & \textbf{33.1} \inc{3.7} & 25.6 & 63.0 & 17.9 & 36.4 & 18.5 & 14.3 & 7.7 & 97.5 & 7.7 & 10.5 & 4.2 & 0.0 \\

            \midrule

            \multicolumn{15}{l}{\textbf{GUIRepair vs. Ours (GPT-5.2)}} \\
            GUIRepair & GPT-5.2 & 29.8 & 28.2 & 61.1 & 11.9 & 27.3 & 13.0 & 19.0 & 0.0 & 97.5 & 15.4 & 0.0 & 4.2 & 0.0 \\

            \rowcolor{gray!10}
            \textbf{FailureMem (Ours)} & GPT-5.2 & \textbf{33.3} \inc{3.5} & 28.2 & 66.7 & 15.7 & 27.3 & 20.4 & 9.5 & 10.3 & 97.5 & 15.4 & 10.5 & 4.2 & 0.0 \\

            \midrule

            \multicolumn{15}{l}{
            \textbf{GUIRepair vs. Ours (Claude 4 Sonnet )}} \\
            GUIRepair & Claude 4 & 28.6 & 20.5 & 61.1 & 11.9 & 45.5 & 11.1 & 4.8 & 5.1 & 96.2 & 7.7 & 0.0 & 0.0 & 0.0 \\
            \rowcolor{gray!10}
            \textbf{FailureMem (Ours)} & Claude 4 & \textbf{32.5} \inc{3.9} & 23.1 & 68.5 & 14.9 & 36.4 & 16.7 & 9.5 & 7.7 & 97.5 & 15.4 & 13.2 & 4.2 & 0.0 \\

            \midrule

            \multicolumn{15}{l}{
            \textbf{GUIRepair vs. Ours (Claude 4.5 Opus thinking)}} \\
            GUIRepair & Claude 4.5 & 31.5 & 23.1 & 64.8 & 13.4 & 54.5 & 16.7 & 9.5 & 5.1 & 97.5 & 15.4 & 5.3 & 4.2 & 0.0 \\
            \rowcolor{gray!10}
            \textbf{FailureMem (Ours)} & Claude 4.5 & \textbf{33.8} \inc{2.3} & 23.1 & 68.5 & 17.2 & 36.4 & 18.5 & 19.0 & 7.7 & 97.5 & 15.4 & 13.2 & 4.2 & 0.0 \\

            \bottomrule
        \end{tabular}
    }
     \caption{\textbf{Comprehensive Performance on SWE-bench Multimodal.} We compare FailureMem against state-of-the-art baselines. Columns under \textit{Repository Breakdown} show the resolved rate (\%) for each specific repository.}
    \label{tab:main_results}
    \vspace{-2mm}
\end{table*}

\subsection{Experimental Setup}
\vspace{-2mm}
\paragraph{Benchmark.} We evaluate FailureMem on SWE-bench Multimodal (SWE-bench M)~\cite{SWEBenchM}, which consists of 617 real-world GitHub issues spanning 17 popular JavaScript repositories. This benchmark requires multimodal reasoning, as visual information is strictly necessary for resolving over 83\% of the tasks~\cite{SWEBenchM}.
\vspace{-2mm}
\paragraph{Baselines.} We primarily benchmark against GUIRepair~\cite{guirepair}, the current SOTA open-source multimodal workflow-based approach. We also report results from text-centric workflows like Agentless~\cite{LingmaAgent} and generalist agents like SWE-agent~\cite{SWE-agent}, alongside commercial systems (e.g., Globant Code Fixer) as upper-bound references.
\vspace{-2mm}
\paragraph{Implementation Details.} We reproduce the official implementation of GUIRepair. All experiments use a sampling temperature of 0 for deterministic generation and follow the Pass@1 evaluation protocol, allowing only a single predicted patch without re-ranking. Final results are reported as the average of three independent runs. Experiments are conducted on four NVIDIA A100 (80GB) GPUs.

\subsection{Results and Discussion}
Table~\ref{tab:main_results} shows FailureMem consistently outperforms the baseline GUIRepair across all tested foundation models. Specifically, when equipped with GPT-5.1, our framework achieves a resolved rate of 33.1\%, yielding a significant absolute improvement of 3.7\% over the baseline. We observe similar consistent gains with GPT-4.1 (+2.3\%) and Claude 4.5 (+2.3\%), validating that the benefits of our failure-aware mechanism are model-agnostic. Furthermore, FailureMem surpasses all other reference baselines, including both agent-based frameworks such as SWE-agent and workflow-based systems like Agentless Lite.
\vspace{-2mm}
\paragraph{Ablations.}
\label{sec:ablation}
To analyze the contributions of each component, we perform an ablation study on SWE-bench Multimodal with the GPT-5.1 backend (See Table~\ref{tab:ablation}). We define a \textbf{Base} configuration that mirrors the GUIRepair workflow, a standard multimodal agent without dynamic exploration or historical memory. 


The \textbf{+Active Perception} variant yields a 1.4\% improvement over the baseline (28.6\% $\to$ 30.0\%). Our error analysis indicates that the Base agent frequently hallucinates DOM element attributes when processing downsampled full-page screenshots. The inclusion of the Crop and Grounding tools effectively mitigates this by allowing the model to request high-resolution re-sampling of specific regions, thereby converting vague visual signals into precise coordinate-level constraints.


The \textbf{+Bash} environment contributes a 1.6\% gain (30.2\%). We observed that purely generative agents often propose theoretically valid but environmentally incompatible patches, such as referencing non-existent relative paths or deprecated API versions. The interactive shell shifts the paradigm from ``write-and-pray'' to ``verify-then-commit'', enabling the agent to validate file structures and dependencies before attempting a fix.


Notably, \textbf{+FailureMem} delivers the largest individual performance boost (+2.3\%, reaching 30.9\%). This suggests that a significant portion of repair failures stems not from a lack of capability, but from \textit{cognitive recurrence}—the tendency to repeatedly attempt plausible but incorrect solutions (e.g., modifying the view layer for a state logic bug). By injecting negative constraints, FailureMem effectively prunes these high-probability failure branches early in the reasoning process.

\begin{table}[!th]
    \centering
    
    \resizebox{\columnwidth}{!}{
        \begin{tabular}{lcc}
            \toprule
            \textbf{Configuration} & \textbf{Resolved Rate (\%)} & \textbf{Improvement} \\
            \midrule
            \textbf{Base} (Standard Workflow) & 28.6 & - \\
            \midrule
            + Active Perception & 30.0 & \inc{1.4} \\
            + Bash Environment & 30.2 & \inc{1.6} \\
            + FailureMem (Memory Only) & 30.9 & \inc{2.3} \\
            \midrule
            \textbf{Full Framework} & \textbf{33.1} & \textbf{\inc{4.5}} \\
            \bottomrule
        \end{tabular}
    }
    \caption{\textbf{Ablation Analysis on GPT-5.1.} We compare the Full Framework against the Base and single-component variants. ``Active Perception'' enables Crop/Grounding tools; ``Bash'' provides an interactive shell; ``FailureMem'' injects negative memory constraints.}
    \label{tab:ablation}
    \vspace{-2mm}
\end{table}


The \textbf{Full Framework} achieves 33.1\%, a cumulative improvement of +4.5\%. This performance exceeds any single component, confirming that the modules address orthogonal failure modes: Active Perception ensures the \textit{input} is accurate, FailureMem ensures the \textit{plan} is sound, and Bash ensures the \textit{execution} is valid.

\vspace{-2mm}
\paragraph{Memory Ablations.}
\label{sec:memory_ablation}
We investigate the distinct contributions of the cognitive and code layers within FailureMem. To rigorously isolate these components, we conducted experiments using the GPT-5.1 backbone. In all experimental settings, the retrieval mechanism remains constant: the Selector Agent identifies the top-3 relevant memory entries based on the Contextual Layer. We vary only the specific information fields injected into the repair agent's context. Table~\ref{tab:mem_ablation} presents the results of five configurations. Variant (A) retains only the cognitive components (Diagnosis and Negative Constraints). Variant (B) simulates a standard RAG baseline by providing only the Golden Patch Summary. Variant (C) represents a positive-reinforcement setup, providing both the Golden Principle and Golden Patch Summary. Variant (D) replaces the natural language summaries in the Code Layer with raw git diffs. Finally, Configuration (E) represents our full framework.

\begin{table}[t]
    \centering
  
    \resizebox{\columnwidth}{!}{
        \setlength{\tabcolsep}{3.5pt}
        \begin{tabular}{l ccc c c}
            \toprule
            \multirow{2}{*}{\textbf{Configuration}} & \multicolumn{3}{c}{\textbf{Memory Fields}} & \multicolumn{2}{c}{\textbf{Performance}} \\
            \cmidrule(lr){2-4} \cmidrule(l){5-6}
             & \textbf{Cognitive} & \textbf{Code} & \textbf{Failure} & \textbf{Resolved} & \textbf{Rate (\%)} \\
            \midrule
            (A) Cognitive Only & \checkmark & - & \checkmark & 154 / 517 & 29.8 \\
            (B) Code Only (RAG) & - & \checkmark & - & 158 / 517 & 30.6 \\
            (C) Positive Only & \checkmark & \checkmark & - & 158 / 517 & 30.6 \\
            (D) Raw Patch & \checkmark & \text{Diffs} & \checkmark & 159 / 517 & 30.8 \\
            \rowcolor{gray!10}
            \textbf{(E) Full Summary (Ours)} & \checkmark & \checkmark & \checkmark & \textbf{171 / 517} & \textbf{33.1} \\
            \bottomrule
        \end{tabular}
    }
      \caption{\textbf{Component Analysis of FailureMem.} We evaluate the impact of different memory compositions. The \textit{Failure} column indicates the inclusion of Failed Patch Summaries and Negative Constraints. Configuration (E) represents the proposed method.}
    \label{tab:mem_ablation}
    \vspace{-2mm} 
\end{table}

Variant (A) relies on high-level reasoning, providing the Cognitive Diagnosis and Negative Constraints without concrete code examples. This configuration yields the lowest performance at 29.8\%. While the model correctly identifies architectural constraints, it struggles to translate abstract directives into valid syntax specific to the target repository. In multimodal repair tasks involving complex DOM APIs, abstract reasoning must be supported by explicit code references provided by the Code Layer to facilitate correct implementation.

We further analyzed whether the form of the Code Layer influences performance. Variant (D) replaces the distilled patch summaries with raw code diffs, resulting in a performance drop to 30.8\%. Raw patches frequently contain project-specific artifacts, such as variable naming conventions or unrelated context lines, which introduce noise into the context window. Natural language summaries extract the core repair logic from these implementation details. This filtering process allows the agent to transfer the underlying fix pattern to the current issue more effectively than direct code copying.

The comparison between Variant (C) and our proposed framework (E) highlights the core contribution of this work. Variant (C) provides comprehensive positive guidance, including both Golden Principles and Golden Patch Summaries, yet its performance plateaus at 30.6\%, identical to the Code Only baseline. This result suggests that providing correct examples alone is insufficient to prevent the model from repeating common mistakes. The inclusion of failure components in Configuration (E) improves the resolved rate to 33.1\%. By explicitly contrasting the Failed Patch Summary with the Golden Patch Summary, the framework provides a discriminative signal. This signal enables the agent to distinguish between the correct solution and plausible but incorrect alternatives that positive examples alone cannot identify.

\begin{figure}[!th]
    \centering
    \includegraphics[width=0.4\textwidth]{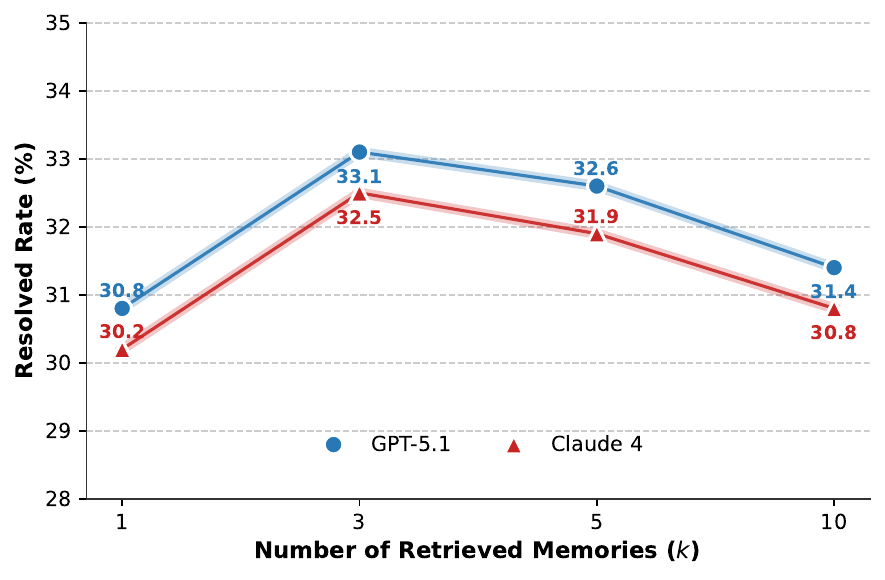}
    \vspace{-10pt}
    \caption{\textbf{Impact of Retrieval Size $k$.} Both models exhibit an inverted U-shape trend, peaking at $k=3$. 
    }
    \label{fig:k_sensitivity}
    \vspace{-2mm}
\end{figure}

\vspace{-2mm}
\paragraph{Memory Size Impacts.}
\label{sec:k_analysis}
We evaluate different numbers of retrieved memory entries, $k \in \{1,3,5,10\}$, using GPT-5.1 and Claude 4. In Figure~\ref{fig:k_sensitivity}, both models follow an inverted U-shaped trend: performance improves from $k=1$ to $k=3$, peaking at 33.1\% (GPT-5.1) and 32.5\% (Claude 4), then declines as $k$ increases further. At $k=10$, resolved rates drop to 31.4\% and 30.8\%, likely due to context dilution from excessive retrieved code. We therefore set $k=3$ as the default to balance information coverage and reasoning focus.

\section{Related Work}
Research on LLM-based Automated Program Repair (APR) has progressed from early fine-tuning approaches~\cite{RAPGen,FitRepair,MoRepair,RepairLlama} and prompting-based methods~\cite{LLM4APR_Fan,LLM4APR_Xia,AlphaRepair,GAMMA,ChatRepair,DRCodePilot,xu2025aligning,CREF,PredicateFix,TypeFix,SRepair, jiang2025screencoderadvancingvisualtocodegeneration, fan2026exploringreasoningrewardmodel}, to autonomous agents for repository-level issue resolution~\cite{SWEBench,Moatless,Marscode,APRAgent_Study}. While systems such as SWE-agent~\cite{SWE-agent} and Agentless~\cite{LingmaAgent} improve general debugging, they lack the multimodal capabilities required for GUI-based tasks. Specialized tools exist for limited visual domains, such as UI design~\cite{yuan2025designrepair} and accessibility~\cite{Iris}, but general-purpose multimodal repair remains underexplored. The current SOTA, GUIRepair~\cite{guirepair}, extends agentless workflows with cross-modal reasoning to incorporate visual information. However, existing approaches suffer from passive perception, relying only on static inputs without active visual exploration, and statelessness, treating each repair attempt independently without leveraging historical experience. These limitations motivate the need for agents with active tool use and memory mechanisms to support more robust multimodal reasoning.

\section{Conclusion}
We propose FailureMem, an experience-driven framework for Multimodal Automated Program Repair that integrates a hybrid workflow–agent architecture, active visual grounding, and a hierarchical failure memory bank. Experiments on SWE-bench Multimodal show consistent improvements over strong baselines.


\section*{Limitations}

Despite FailureMem's performance gains, two limitations remain. The iterative agentic loop and detailed memory contexts increase inference costs to \$0.33 per issue, a 13\% rise compared to GUIRepair's \$0.29. We consider this tradeoff acceptable given the 3.7\% improvement in resolution rate and the relatively low cost of autonomous compute versus human effort. Additionally, FailureMem depends on the diversity of its offline memory bank. Rare or unprecedented failure modes without historical analogues bypass the contrastive distillation process, causing the system to revert to standard agentic behavior.

\section*{Ethics Statement}
This work studies Multimodal Automated Program Repair and proposes FailureMem to improve the reliability of automated software debugging. Our goal is to assist developers by identifying and repairing software defects more effectively. The system is intended as a decision-support tool, and generated patches should be reviewed and validated by developers before deployment in production environments. Automated repair systems may occasionally generate incorrect or incomplete patches, and human oversight remains essential for ensuring software safety and correctness.

The memory bank used in FailureMem is constructed from publicly available software repositories and issue reports, which may include textual descriptions and screenshots of user interfaces. These artifacts are processed to extract abstract repair knowledge rather than storing raw sensitive content. We do not intentionally collect personal or confidential data, and the framework is designed to store summarized repair experiences rather than proprietary code whenever possible. Nevertheless, practitioners should ensure that memory construction follows appropriate data governance policies when applied to private repositories.

Automated program repair technologies may also be misused if deployed without sufficient safeguards. For example, blindly applying automatically generated patches could introduce regressions or security vulnerabilities. Our framework therefore emphasizes learning from failed repair attempts and negative constraints, aiming to reduce recurring mistakes rather than encouraging fully autonomous code modification.

Finally, the use of large language models and multimodal reasoning systems requires substantial computational resources. Our experiments are conducted on GPU infrastructure, which has environmental implications. Future work will explore more efficient memory retrieval and reasoning strategies to reduce computational costs while maintaining repair effectiveness.

\bibliography{custom}

\appendix
\section{Data Construction for FailureMem}
\label{sec:app_data_construction}

We detail the offline construction process of the Failure Memory Bank, specifically focusing on how empirical failure trajectories are collected and distilled into structured memory entries.

\subsection{Data Source and Trajectory Collection}
\label{sec:app_data_source}
Our memory bank is strictly derived from the development set of the SWE-bench Multimodal benchmark, which comprises 102 real-world visual software issues. To capture natural, high-probability failure patterns, we first ran FailureMem without the Failure Memory Bank (i.e., using only the hybrid workflow-agent architecture and active perception tools, powered by GPT-5.1) to attempt all 102 instances.

The evaluation yielded a clear behavioral split: the memory-free variant correctly resolved 18 issues but failed to produce valid patches for the remaining 84 instances. We exclusively utilize these 84 failed trajectories as our source material for memory construction. The 18 successful instances were intentionally excluded to ensure the memory bank remains focused on learning from failures rather than redundant positive imitation. For each of the 84 failed instances, we collected the incorrect patch output ($P_{fail}$) from this variant to serve as the contrastive counterpart to the developer-verified ground truth ($P_{gold}$).

\subsection{Offline Distillation Pipeline}
\label{sec:app_distillation_pipeline}
To transform raw failed trajectories into the hierarchical memory entries defined in Section~\ref{sec:failuremem}, we implemented an automated offline distillation pipeline. We utilized Gemini 3 Pro as the reasoning engine to conduct a contrastive root cause analysis.

For each of the 84 failed instances, the pipeline constructs a multimodal context tuple $\langle \mathcal{D}, \mathcal{I}, P_{fail}, P_{gold} \rangle$, where:
\begin{itemize}
    \item $\mathcal{D}$: The original natural language issue description.
    \item $\mathcal{I}$: Up to three visual symptom screenshots (encoded as base64 images).
    \item $P_{fail}$: The incorrect patch generated by the memory-free variant.
    \item $P_{gold}$: The correct developer patch.
\end{itemize}

The distillation model is prompted to compare $P_{fail}$ against $P_{gold}$ and abstract the multimodal inputs into our three-layer memory architecture:

\textbf{Contextual Layer ($\mathcal{L}_{ctx}$):} The model synthesizes the raw inputs into two text-based fields: the Issue Summary, which abstracts the bug scenario, and the Visual Analysis, which textually describes the visual symptoms (e.g., ``The modal overlay obscures the navigation bar''). This deliberate translation from raw screenshots to text minimizes token consumption and prevents visual background noise from distracting the selector agent during the retrieval phase.

\textbf{Cognitive Layer ($\mathcal{L}_{cog}$):} The model extracts high-level reasoning guidance consisting of three components. It generates a Cognitive Diagnosis that explains the causality behind the failure; formulates a Negative Constraint that explicitly forbids the specific incorrect strategies (e.g., ``Do not modify downstream rendering logic for upstream data errors''); and outlines a Golden Principle that defines the correct design pattern.

\textbf{Code Layer ($\mathcal{L}_{code}$):} To provide concrete references, the model generates a Failed Patch Summary and a Golden Patch Summary. These summaries highlight the specific implementation divergence between the incorrect and correct solutions, offering the agent an empirical reference for the repair strategy while filtering out project-specific noise found in raw diffs.

To ensure the reliability of the memory bank, the pipeline includes a strict field-level validation step. Any generated entry missing the required structural layers, containing empty strings for critical reasoning steps, or exhibiting hallucinated syntax is automatically rejected and retried with exponential backoff. This rigorous filtering resulted in a high-fidelity memory bank containing 84 well-structured memory entries ready for retrieval.

\subsection{Memory Bank Statistics and Case Example}
\label{sec:app_memory_example}

The final memory bank comprises 84 distinct memory entries derived from the development set failures. Table~\ref{tab:repo_distribution} details the distribution of these empirical failure trajectories across the repositories included in the SWE-bench Multimodal benchmark. This distribution ensures the retrieved memory entries cover a wide spectrum of visual structures, rendering frameworks (e.g., PDF generation, Canvas drawing), and architectural patterns.

\begin{table}[ht]
    \centering
    \caption{\textbf{Repository Distribution of the Memory Bank.} The 84 failed trajectories used for extracting negative constraints span 5 distinct repositories.}
    \label{tab:repo_distribution}
    \vspace{5pt}
    \resizebox{0.85\columnwidth}{!}{
        \begin{tabular}{lc}
            \toprule
            \textbf{Repository} & \textbf{Count of Failed Trajectories} \\
            \midrule
            Automattic/wp-calypso & 37 \\
            chartjs/Chart.js & 16 \\
            markedjs/marked & 11 \\
            diegomura/react-pdf & 10 \\
            processing/p5.js & 10 \\
            \midrule
            \textbf{Total Memory Entries} & \textbf{84} \\
            \bottomrule
        \end{tabular}
    }
    \vspace{-2mm}
\end{table}

To illustrate the exact payload injected into the agent's context during Phase 3, we present a complete, distilled memory entry for the instance \texttt{Automattic\_\_wp-calypso-21964} in Figure~\ref{fig:memory_example_box}. Extraneous metadata utilized exclusively for offline pipeline routing has been omitted for clarity.

\begin{figure*}[t]
    \centering
    \hrule
    \vspace{0.2cm}
    \begin{minipage}{0.96\textwidth}
    \small 
    
    \textbf{Memory Entry Example: \texttt{Automattic\_\_wp-calypso-21964}}
    \vspace{0.1cm}
    \hrule
    \vspace{0.2cm}

    \textbf{Contextual Layer ($\mathcal{L}_{ctx}$)}
    \begin{itemize}
        \item \textbf{Issue Summary:} OAuth \texttt{client\_id} parameter is dropped when navigating from Signup back to Login form, causing loss of custom branding/styling.
        \item \textbf{Visual Analysis:} The login page reverts to generic styling instead of custom branded styling after a user clicks 'Already have an account?'. The expected behavior is that the login page should retain custom branding (driven by \texttt{client\_id}) when returning from the Signup form.
    \end{itemize}

    \vspace{0.1cm}
    \textbf{Cognitive Layer ($\mathcal{L}_{cog}$)}
    \begin{itemize}
        \item \textbf{Cognitive Diagnosis:} The agent attempted to fix a View-layer link generation issue by modifying Controller-layer route initialization, failing to propagate data to the actual UI component. It calculated \texttt{initialUrl} in \texttt{client/signup/controller.js} to include query strings, assuming this would implicitly preserve the parameters during navigation. However, the 'Back to Login' link is explicitly rendered by a React component using a helper function, so the controller-level variable had no effect on the rendered UI. The Golden Patch demonstrates the 'Selector Pattern': retrieving the persisted state from the store within the View component and passing it explicitly to the URL generation utility.
        \item \textbf{Negative Constraints:}
        \begin{itemize}
            \item[--] Do NOT attempt to fix UI link generation issues by modifying controller initialization variables that are not passed to the view.
            \item[--] Do NOT assume URL parameters persist automatically across route changes; explicitly inject them into link generators.
        \end{itemize}
        \item \textbf{Golden Principle:} Reconstruct navigation state explicitly in View components using Store Selectors, rather than relying on implicit Controller context preservation.
    \end{itemize}

    \vspace{0.1cm}
    \textbf{Code Layer ($\mathcal{L}_{code}$)}
    \begin{itemize}
        \item \textbf{Failed Patch Summary:} Modified \texttt{client/signup/controller.js} (Controller). Added logic to reconstruct \texttt{initialUrl} from \texttt{context} including query strings. The bug wasn't about the controller losing context, but about the View component (\texttt{SignupForm}) generating a link that lacked the parameter. Modifying a local variable in the controller without passing it to the view has no effect on the rendered HTML.
        \item \textbf{Golden Patch Summary:} The fix involves updating the URL generation utility to support the \texttt{client\_id} parameter, and connecting the View component to the Redux store to retrieve the current client ID.
        \begin{itemize}
            \item[--] \texttt{lib/paths/login/index.js} (Utility): Updated the \texttt{login} function to append \texttt{?client\_id=...} if \texttt{oauth2ClientId} is provided, centralizing URL logic to ensure consistency.
            \item[--] \texttt{components/signup-form/index.jsx} (View): Used the \texttt{connect} wrapper to retrieve \texttt{getCurrentOAuth2Client(state)} from Redux and explicitly injected \texttt{oauth2ClientId} into the \texttt{getLoginLink} method, ensuring the component has access to global context data.
        \end{itemize}
    \end{itemize}
    
    \end{minipage}
    \vspace{0.2cm}
    \hrule
    \caption{\textbf{An instantiated Memory Entry} utilized during the final repair phase. The structured fields align with the defined hierarchical memory architecture ($\mathcal{L}_{ctx}, \mathcal{L}_{cog}, \mathcal{L}_{code}$). It is presented across both columns for readability.}
    \label{fig:memory_example_box}
\end{figure*}

\section{Extended Methodology Details}
\label{appendix:methodology}

This appendix provides detailed engineering descriptions of the core components 
discussed in the main text: the skeleton compression strategy used for context 
management (\S\ref{app:skeleton}), the implementation of Active Perception tools 
(\S\ref{app:perception}), and the sandboxed execution environment for codebase 
exploration (\S\ref{app:bash}).

\subsection{Skeleton Compression Strategy}
\label{app:skeleton}

Providing the complete source code of all candidate files to the model is often 
impractical: a single file in a large front-end repository can span thousands of 
lines, and the full context of multiple files easily exceeds the effective context 
window of current language models. To address this, we design a \emph{skeleton 
compression} strategy that retains the structural outline of each file while 
aggressively removing implementation bodies, yielding a concise yet informative 
representation.

\paragraph{AST-Based Structure Extraction.}
We leverage an Abstract Syntax Tree (AST) parser to identify the structural 
elements of each source file. Concretely, we employ the Babel parser---a 
widely-used, plugin-extensible parser for JavaScript and TypeScript---invoked as 
an external process. The parser is configured with a comprehensive set of syntax 
plugins (including JSX, TypeScript, class properties, optional chaining, 
nullish coalescing, decorators, pipeline operators, and others) to ensure broad 
compatibility across diverse repository coding conventions. For repositories 
that use non-standard or unconventional coding patterns, we maintain 
repository-specific parser configurations that handle edge cases such as 
recursive node traversal with cycle detection or null-safe property access 
during AST walking.

The parser performs a recursive traversal of the AST and extracts two categories 
of structural elements:

\begin{itemize}[leftmargin=*,nosep]
    \item \textbf{Classes}: Each class declaration is recorded with its name, 
    start line, end line, and a list of its methods (each with name, start line, 
    and end line). Both named and anonymous class declarations are captured, 
    including those assigned via \texttt{module.exports}.
    \item \textbf{Functions}: Named function declarations, arrow function 
    expressions, and function expressions assigned to variables or object 
    properties are recorded with their names and line spans.
\end{itemize}

\paragraph{Skeleton Construction.}
Given the extracted structure, we construct the skeleton representation through 
the following procedure:

\begin{enumerate}[leftmargin=*,nosep]
    \item \textbf{Initialize an empty template} of the same length as the 
    original file (one empty string per line).
    \item \textbf{Retain class boundaries}: For each class, copy the 
    declaration header (start line) and the closing brace (end line) into the 
    template. For each method within the class, similarly retain the method 
    signature line and closing line.
    \item \textbf{Retain function signatures}: For each function, copy the 
    start line and the end line. Additionally, if the function signature spans 
    multiple lines (e.g., functions with long parameter lists), we expand 
    downward from the start line for up to 20 additional lines, stopping when 
    we encounter an empty line or a line ending with a block delimiter 
    (\texttt{\{} or \texttt{\}}). This ensures that multi-line function 
    signatures are preserved in their entirety.
    \item \textbf{Preserve comments, imports, and exports}: All comment lines 
    (single-line \texttt{//}, multi-line \texttt{/* */}, and JSDoc \texttt{*} 
    lines), \texttt{import} statements, and \texttt{export} statements are 
    unconditionally retained. These provide semantic context about the file's 
    dependencies and public interface.
    \item \textbf{Collapse consecutive blank lines}: Sequences of more than 
    two consecutive blank lines are collapsed to at most two, preventing large 
    gaps where function bodies were removed.
\end{enumerate}

The compression is applied conditionally: files shorter than a configurable 
threshold (in lines) are provided in their entirety, since the overhead of 
full inclusion is minimal. Files exceeding this threshold are compressed. In 
rare cases where the compressed skeleton itself exceeds 5{,}000 lines (e.g., 
for exceptionally large auto-generated files), we truncate to the first 
5{,}000 lines. If compression yields an empty result (e.g., for declarative 
configuration files with no class or function structure), the full file 
content is used as a fallback.

This skeleton format typically achieves a 5--20$\times$ compression ratio 
compared to the original file, making it feasible to present dozens of 
candidate files simultaneously within a single prompt while preserving the 
structural information necessary for fault localization.

\subsection{Implementation of Active Perception Tools}
\label{app:perception}

Our framework equips the agent with two complementary visual analysis 
tools---\textsc{Crop} and \textsc{Grounding}---that enable it to actively 
manipulate and inspect bug scenario screenshots during the repair process. 
Both tools operate on raw pixel coordinates specified directly by the agent.

\paragraph{Coordinate Specification.}
When the agent invokes a visual tool, it outputs a bounding box in the form 
$[x_{\min}, y_{\min}, x_{\max}, y_{\max}]$, where coordinates are specified 
in absolute pixel units relative to the top-left corner of the image. The agent 
determines these coordinates by reasoning about the spatial layout of the 
screenshot based on its visual understanding. No external grounding module 
(e.g., Set-of-Mark prompting or DOM tree parsing) is employed; the agent 
directly estimates pixel regions from the rendered screenshot. An image index 
parameter allows the agent to select which screenshot to operate on when 
multiple bug scenario images are available.

\paragraph{Crop Tool.}
The \textsc{Crop} tool extracts a sub-region from a screenshot to enable 
detailed inspection of fine-grained visual artifacts. Upon receiving the 
agent's bounding box, the system:
\begin{enumerate}[leftmargin=*,nosep]
    \item Decodes the base64-encoded screenshot into a pixel buffer.
    \item Validates and clamps the bounding box coordinates to the image 
    boundaries to prevent out-of-bounds errors.
    \item Extracts the specified rectangular region.
    \item Re-encodes the cropped region as a new image and injects it into 
    the subsequent conversation turn as a user message containing the cropped 
    image.
\end{enumerate}
This tool is designed for scenarios requiring pixel-level examination, such 
as verifying whether a border is 1\,px too thick, whether an icon is rendered 
at incorrect resolution, or whether text overflow is occurring at a specific 
breakpoint.

\paragraph{Grounding Tool.}
The \textsc{Grounding} tool annotates the original screenshot with a 
bounding box overlay to highlight the bug-affected region while preserving 
the full page context. Given the agent's bounding box and an optional text 
label, the system:
\begin{enumerate}[leftmargin=*,nosep]
    \item Decodes the screenshot into a pixel buffer.
    \item Draws a colored rectangle (red, 3\,px width) at the specified 
    coordinates on a copy of the image.
    \item If a text label is provided, renders it above the bounding box 
    with a contrasting background for readability.
    \item Re-encodes the annotated image and injects it into the 
    conversation as a new visual input.
\end{enumerate}
This tool is particularly useful for layout and positioning bugs, where 
the spatial relationship between the highlighted region and the 
surrounding page elements is critical for diagnosis.

\paragraph{Tool Usage Protocol.}
Both tools follow a strict single-tool-per-turn protocol: the agent 
issues exactly one tool call per response, then waits for the system to 
return the actual result (the processed image) before proceeding. This 
prevents hallucinated tool outputs and ensures that subsequent reasoning 
is grounded in real visual evidence. The agent is encouraged to use 
visual tools as its \emph{first} action when bug scenario images are 
available, establishing a concrete visual understanding before 
proceeding to code-level analysis. The processed images (both 
originals and tool outputs) are persisted to disk for post-hoc analysis.

\subsection{Bash Environment Constraints}
\label{app:bash}

To support codebase exploration during patch generation, we provide the 
agent with a sandboxed shell execution environment. This allows the 
agent to inspect files, search for patterns, and understand the 
repository structure when the provided code snippets are insufficient 
for generating a correct fix.

\paragraph{Execution Model.}
Each shell command is executed as an independent subprocess with its 
working directory set to the repository root. The \texttt{cd} command 
is explicitly blocked, as directory changes do not persist across 
invocations; instead, the agent uses relative paths from the project 
root (e.g., \texttt{cat src/components/Button.js}). This stateless 
execution model simplifies security enforcement and prevents 
path-related confusion.

\paragraph{Security Isolation.}
We enforce a multi-layered security policy to ensure that the agent's 
shell access is strictly read-only:

\begin{itemize}[leftmargin=*,nosep]
    \item \textbf{Command blacklist}: Destructive file system operations 
    (\texttt{rm}, \texttt{mv}, \texttt{cp}, \texttt{mkdir}, 
    \texttt{touch}, \texttt{chmod}), version control mutations 
    (\texttt{git reset}, \texttt{git checkout}, \texttt{git clean}, 
    \texttt{git merge}, \texttt{git rebase}, \texttt{git push}), 
    network utilities (\texttt{curl}, \texttt{wget}, \texttt{ssh}, 
    \texttt{scp}), package managers (\texttt{npm}, \texttt{pip}, 
    \texttt{apt}), and process control commands (\texttt{kill}, 
    \texttt{sudo}) are rejected before execution.
    \item \textbf{Redirect blocking}: Output redirection operators 
    (\texttt{>}, \texttt{>>}, \texttt{2>}, \texttt{\&>}) are 
    forbidden to prevent any file writes. Pipe operators (\texttt{|}) 
    are allowed for command chaining (e.g., \texttt{grep | head}) but 
    are validated to ensure the downstream command is not a 
    write-capable utility.
    \item \textbf{Injection prevention}: Command substitution 
    (\texttt{\$(...)} and backticks) and background execution 
    (\texttt{\&}) are blocked to prevent privilege escalation and 
    uncontrolled process spawning. In-place file editing via 
    \texttt{sed -i} is specifically detected and rejected while 
    allowing read-only \texttt{sed} usage for text extraction.
    \item \textbf{Command chaining}: Semicolon-separated command 
    chains are permitted, but each sub-command is individually 
    validated against the blacklist before execution.
\end{itemize}

\paragraph{Resource Limits.}
To prevent runaway processes and excessive context injection, we impose 
the following constraints:

\begin{itemize}[leftmargin=*,nosep]
    \item \textbf{Timeout}: Each command is subject to a 120-second 
    execution timeout. Commands exceeding this limit are terminated, 
    and the agent receives a timeout notification.
    \item \textbf{Output truncation}: Command output is capped at 300 
    lines or 50\,KB (whichever is reached first). When truncation 
    occurs, the agent is informed and advised to narrow its query 
    (e.g., by using more specific patterns or limiting the search 
    scope to a subdirectory).
    \item \textbf{Directory exclusion}: Common non-source directories 
    (\texttt{node\_modules}, \texttt{.git}, \texttt{dist}, 
    \texttt{build}, \texttt{coverage}, etc.) are excluded from 
    search and traversal operations to reduce noise and improve 
    response time.
\end{itemize}

\paragraph{Permitted Operations.}
Within these constraints, the agent can freely execute read-only 
operations including file reading (\texttt{cat}, \texttt{head}, 
\texttt{tail}), pattern searching (\texttt{grep} with recursive, 
case-insensitive, and context-line options), file discovery 
(\texttt{find}, \texttt{ls}), and text processing (\texttt{wc}, 
\texttt{sort}, \texttt{awk} in read-only mode). This set of 
operations is sufficient for the agent to navigate unfamiliar 
codebases, trace import chains, locate related files, and 
gather the contextual information necessary for patch generation.

\end{document}